\documentstyle[12pt]{article}
\begin{document}
\topskip 2cm
\begin{titlepage}

\begin{center}
{\large{\bf SUPERNOVA 1987A - TEN YEARS AFTER}}\\ 

\vspace{2.5cm}
{\large Arnon Dar} \\
\vspace{.5cm}
{\sl Department of Physics and Space Research Institute}\\ 
{\sl Technion - Israel Institute of Technology}\\
{\sl Haifa 32000), Israel}\\
\vspace{2.5cm}
\vfil

\begin{abstract}

{\bf Supernova 1987A became a milestone in physics and astronomy. The 
most important things that have been learned from it, the most important
problems yet to be solved and the prospects for learning important new
physics from future observations of nearby supernova explosions are
shortly summarized.}

\end{abstract}

\end{center}
\end{titlepage}
\section{Introduction}

SN1987A, the supernova explosion on February 23, 1987 in the nearby Large
Magellanic Cloud only about $50~kpc$ away, was the brightest supernova
seen since the invention of the telescope. It is the first supernova which
has been visible to the unaided eye since Kepler saw SN1604, the last
Supernova seen in our Milky Way galaxy. It has offered a unique
opportunity to observe for the first time a supernova explosion from a
relatively close distance within the range of various detection
techniques. The first signals that were recorded on Earth were neutrino
signals in the Mont Blanc (Aglietta et al. 1987), Kamiokande (Hirata et
al. 1987), IMB (Bionta et al. 1987) and Baksan (Alexeyev et al. 1988) 
underground detectors and an unconfirmed gravitational wave signal in 
the Rome detector (Amaldi et al. 1987).
They were followed by a spectacular optical flash that began a
few hours later, but was the first signal from 1987A that had been noticed
(McNaught 1987). Observations of SN1987A have continued since then, from
the ground (optical telescopes, radio telescopes, gravitational wave
antenas, high energy $\gamma$ ray Cerenkov telescopes and extensive air
shower arrays) from underground (neutrino telescopes), from high in the
air (detectors aboard high altitude planes and balloons) and from space
(Hubble Space Telescope, X-ray telescopes and $\gamma$- ray telescopes).
They have yielded rich information which is of fundamental importance for
astrophysics as well as for other branches of physics and which is
documented in hundreds of papers and many excellent reviews that have been
published in the scientific literature. I will not attempt to review this
vast literature but rather focus on what I think are the most important
consequences of SN1987A, the most important things that we have learned
from it, the most important problems yet to be solved and the prospects
for learning important new physics from future observations of nearby
supernova explosions. 

\section{The Birth of Extrasolar Neutrino Astronomy} 
Perhaps the most important consequence of SN1987A is the birth of
extrasolar neutrino astronomy: When the first large underground water
Cerenkov detectors, IMB and Kamiokande, were constructed for looking for
proton decays, it was suggested that they can also perform as neutrino
telescopes (e.g., Dar 1983 and references therein) which may detect
neutrino bursts from galactic supernova explosions and the diffuse
cosmological neutrino background from stellar evolution and past
supernovae (e.g., Dar 1985).  This was dramatically demonstrated when the
Kamiokande and IMB telescopes detected the neutrino burst from SN1987A. 
This monumental success has probably convinced physicists and funding
agencies that galactic and extragalactic neutrino astronomy are not just a
dream but are important achievable scientific goals. This has already
resulted in the construction of Superkamiokande, an amazing galactic and
near galactic neutrino telescope. Together with the pioneering studies of
the DUMAND project, SN1987A perhaps also led to the construction of the 
AMANDA experiment under the south pole, the Baikal experiment under lake 
Baikal and to the planned NESTOR and ANTARES deep sea projects in the
Mediterranean sea offshore Pylos in Greece and offshore France,
respectively. The Universe is opaque to very high energy gamma rays
because of electron-positron pair production on intergalactic background
photons. It is, however, transparent to neutrinos. It is anticipated that
when the above experiments will be scaled up to a $1~km^3$, they may
detect very high energy neutrinos from Active Galactic Nuclei at
cosmological distances, from the mysterious Gamma Ray Bursters and from
other unexpected sources. They also may point at the nature and identity
of the cosmic accelerators and help solve the 85 years mystery of the
origin of high energy cosmic rays.  These, to my mind, may be the most
important consequences of SN1987A ... 

\section{Supernova Theory}

Already before SN1987A, the theory of type II supernova explosions (SNeII)
was able to explain many of the observed properties of SNeII that occur at
cosmological distances at a rate of about 1 per second per Universe, but
was not able to explain the exact explosion mechanism (see, e.g., Shapiro
and Teukolsky 1983 and references therein, Bruenn 1987 and references
therein). This has not been changed by SN1987A in spite of continuous
theoretical progress, impressive numerical efforts and many important
refinements in the theory of SNeII as a result of the detailed
observations of both SN1987A and other nearby SNeII. It is now generally
believed that spherical symmetric one-dimensional (1-D) codes with the
best available physics (improved progenitor profiles, improved equation of
state, improved opacities and neutrino transport and general relativistic
effects) cannot reproduce SNeII. 

Let me first summarize the SNeII theory prior to SN1987A, its spectacular
success and its serious problems.

Standard stellar evolution theory predicts that massive stars
$8M_\odot\leq M\leq 20M_\odot$ evolve for $\sim10^7~y$ by the
thermonuclear burning of heavier and heavier fuels and terminate in anion
like {\bf red supergiant} with a white dwarf like central core consisting
primarily of iron group nuclei and supported primarily by electron
degeneracy pressure. When the core mass exceeds the Chandrasekhar mass of
about $1.4M_\odot$, gravity overcomes the degeneracy pressure and collapse
begins (e.g., Arnett 1977; Barkat 1977). The central density of the core
increases quickly and reaches a value where electrons from the top of the
Fermi sea can be captured and convert protons, free and in bound nuclei,
into neutrons via $e^-+p\rightarrow n+\nu_e$. The capture of electrons
results in a short neutronization burst (ms) which stops because of Pauli
blocking by neutrinos which are trapped in the core (because of neutral
current elastic scattering from nuclei). Electron capture from the top of
the Fermi sea by free protons and iron group nuclei reduces degeneracy
pressure and accelerates the collapse. The collapse becomes essentially a
free fall with a time scale $t\sim 1/\sqrt{G\rho}\sim 50~ms$. When the
central density of the core reaches supranuclear density the repulsive QCD
forces between nucleon constituents (quarks and gluons) of the same color
stop the collapse, the core bounces and drives a strong shock wave that
climbs outside through the infalling layers.  The strong shock supported
by energy transport through convection and neutrinos is believed somehow
to reverse the infall velocity of the layers, to overcome their
gravitational binding and to propel them to the observed expansion
velocities of more than 10000 $km~s^{-1}$ which amounts to a total kinetic
energy of about $10^{51}~erg$. The shock is believed to produce the
spectacular light display of SNeII (Grassberg, Imshennik and Nadyozhin
1971): With a velocity which is a considerable fraction of the velocity of
light it takes the shock a few hours to reach the atmosphere of the
supergiant (typical radius of about $10^{13}cm$). When it reaches the
atmosphere it heats it up to a high temperature which produces a UV flash.
However, the integrated luminosity of SNeII ($\sim 10^{49}erg$) and the
total kinetic energy of the ejected shell ($10^{51}erg)$ are only a tiny
fraction of the released energy. Most of the gravitational binding energy
of the collapsed core $(\sim GM^2/R\sim$ a few $10^{53}~erg)$ which is
released in the collapse is converted into thermal energy of a
protoneutron star, which cools slowly $(\sim 10~s)$ by radiating neutrinos
from its surface (Colgate and White 1966; Wilson et al. 1986 and
references therein, Mayle et al 1987 and references therein). The
protoneutron star is essentially opaque to neutrinos which are thermally
produced mainly via $e^+e^-\rightarrow \nu\bar\nu$ in the hot core
(central temperature $\sim 30~MeV$) and diffuse slowly to the surface of
last scattering (the ``neutrinosphere'') where they are emitted with a
much smaller temperature, typically $3-4~MeV$ for electron neutrinos and
$7-8~MeV$ for $\mu$ and $\tau$ neutrinos, which can be predicted from
quite general considerations (e.g., Dar 1987). 

SN1987A provided a dramatic confirmation of these predictions of the
theory of SNeII. SN1987A was caused by the violent death of a massive star
$(\sim 20M_\odot)$. The integrated light emission ($\sim 10^{49}erg$) and
the kinetic energy of the expanding shell ($\sim 10^{51}erg$) consisted
only of a tiny fraction of the energy released by SN1987A.  Most of the
energy (a few $10^{53}erg $) was radiated in neutrinos, which indeed were
detected by the Mont Blanc, Kamiokande, IMB and Baksan underground
detectors. As expected, neutrino emission preceded the first UV light
flash by a few hours. The average energy of the $\bar\nu_e$'s was $\sim
13~MeV$ (temperature of about $4~ MeV$) and the duration of the neutrino
burst was $\sim 10~s$. This energy is consistent with the gravitational
binding energy released in the formation of a neutron star in stellar core
collapse. The UV flash and its spectral evolution was well fitted by a
shock wave reaching the surface of the supergiant star and heating it. The
detection of $\gamma$-ray lines and infrared emission lines confirmed that
the exponential decay of the supernova light curve was because the remnant
was being heated by radioactivity from isotopes made in the explosion,
$0.07M_\odot$ of $^{56}$Co and $0.003M_\odot$ of $^{57}$Co. 

However, some major predictions of SNeII theory were not very successful and 
many puzzles remain. They include:

a. Why was the progenitor of SN1987A a blue supergiant and not a red
supergiant? 
 
b. How were the triple rings around the remnant of SN1987A (Fig 1.) formed?

c. Why was the explosion aspherical, as evident from the debris of 
SN1987A?

d. What is the explosion mechanism of SNeII?

e. Did SN1987A produce a neutron star and when will it become visible?

f. Did SN1987A produce a black hole ?

g. Did SN1987A bang twice?

h. Did SN1987A emit significant gravitational radiation? 

\noindent
{\bf a. The Progenitor}: For the first time the progenitor of a SNeII has
been clearly identified.  After the fading of the optical flash from
SN1987A, careful measurements have shown that a type B3 blue supergiant,
entry number 202 in the declination band $69^0$ south of the equator in a
catalog of LMC giants compiled by N. Sanduleak, which was at the exact
position of SN1987A, disappeared in the explosion, whereas its two blue
neighbor stars (Star 1 and Star 2 at respectively 2.90 and 1.66 arcseconds
away) survived the explosion.  Astronomers were astonished to find that
the progenitor of SN1987A was a blue supergiant and not a red supergiant
as thought to be the case for most SNeII. Two alternative explanations 
have been proposed: Perhaps a $\sim 20M_\odot $ blue star on the main 
sequence swelled up to become a red supergiant, lost mass through a 
stellar wind then contracted and reheated to become a blue supergiant. 
Another explanation that leads to a blue supergiant is that the 
progenitor formed from the merger of two stars in a binary system.
The prior history of Sanduleak -69$^0$ 202 is probably imprinted
in the circumstellar nebulae around SN1987A and will
be able to test the two models.

\noindent 
{\bf b. The Rings:} The gas surrounding SN1987A was expected to be
illuminated by EUV and X-rays (Chevalier 1988) emitted when the explosion
shock wave reached the envelope of the pre-supernova star. Early images
taken by the Hubble Space Telescope, which was launched in April 1990,
unexpectedly have shown (Wampler et al 1990; Jakobsen et al 1991) that the
light emission from the circumstellar gas around the remnant of SN1987A is
localized in three ring like forms along a common axis which passes
through the remnant of SN1987A (see Fig. 1) and is tilted at roughly
$45^0$.  The inner ring is centered on the remnant, has an approximate
radius of $R\approx 6.1\times 10^{17} cm$ ($0.65~ly$), a mass about 0.2 to
0.4 $M_\odot$ and a radial velocity $v_r\approx 10~km~s^{-1}$. The ring is
also extraordinarily symmetric and highly localized in both space ($\delta
R/R\approx 10\%$) and velocity. VLB radio observations and recent HST
observations have shown that the radius of the glowing debris from SN1987A
is now about 0.1 arcseconds (about 15\% of the distance to the ring) and
the expansion speed has been nearly constant, over the past 10 year
history, i.e., $\approx 0.01$ arcsecond per year or $v_r\approx
2500~km~s^{-1}$. This is much slower than the speediest material observed
back in 1987, which reached 30000 $km~s^{-1}$, but probably was of a small
mass which was slowed down by the circumstellar gas.  Thus, it seems that
the rings were there before SN1987A. Various models have been proposed for
the origin of the rings. The same basic structure is seen with HST in the
Hourglass Nebula, suggesting that some common aspects of mass loss were at
work both in this planetary nebula and in SN1987A. Consequently, it was
suggested that the SN1987A rings formed by the illumination of a
pre-supernova red giant wind that was much thicker at the waist than the
poles resulting in an expected hour-glass shape (Luo and McCray 1991; Wang
and Mazzali 1992; Blondin and Lunqvist 1993; Martin and Arnett 1995). It was
suggested that the glow of the rings is formed by recombination of
electrons and atoms that were ionized by the EUV and X-ray flash from
SN1987A in the case of the inner ring, and by the EUV and X-ray emissions
from a relativistic conical jets in the case of the external rings. Other
models assume that the inner ring is a relic from an accretion disk (McCay
and Lin 1994) or from an excretion disk from which the presupernova star
was born (Chen and Colgate 1996). It is also possible that the inner 
and outer rings are thin flash ionized layers at the inner surfaces
of much greater mass of circumstellar as yet unseen. 

\noindent 
{\bf c. Aspherical Explosion ? Jets?} Recent high resolution VLB radio images
(Gaensler et al 1997) and HST optical images (Pun 1997) of SN1987A and its
inner ring show that the glowing debris of the supernova itself is
elongated along the axis of the rings. It was pointed out that a a merger
of two stars in a binary system (Podsiadlowski 1992) leads to a blue
supergiant progenitor and can explain an equatorial outflow of several
solar masses of gas during a merger of the two stars some 20.000 years
before the explosion. Such a merger would probably yield a progenitor that
is highly flattened by rotation.  If so, the explosion would naturally
blow out preferentially along the polar axis, perhaps even jetting the
ejecta. Although such a model may be plausible, it is not yet well
developed, certainly not universally accepted. If supernovae explode
aspherically it is imprinted upon the ejecta and has additional signatures
such as significant gravitational radiation (M\"onchmeyer et al. 1991),
natal kicks to nascent neutron stars (Burrows and Hayes 1996; Woosley
1987), mixing of iron-peak and r-process nucleosynthetic products,
generation of pulsar magnetic fields and perhaps jetting of the debris. 
 
\noindent
{\bf d. The explosion Mechanism ?}
In spite of impressive theoretical and numerical efforts during the past
ten years, we still do not know how type II supernovae explode and convert
$\sim 1\%$ of their gravitational energy release into kinetic energy of
debris. Since the neutrino observations of SN1987A provided strong support
for the basic picture of SNeII it is widely believed that neutrinos
coupled with convection transport sufficient energy from the core to the
mantle to blow it off.  Because the observed kinetic energy in SNeII is so
steady, many investigators have hoped (and some still do) that
improvements in the input microscopic and macroscopic physics in 
one-dimensional (1-D) spherical symmetric calculations will lead to the
solution. The improvements in microphysics included the use of improved
neutrino opacities at high densities, the inclusion of the neutrino
annihilation $\nu\bar\nu\rightarrow e^+e^-$ mechanism (Goodman, Dar and
Nussinov 1987) and neutrino breamstrahlung $nn\rightarrow nn+\nu\bar\nu$
(Suzuki 1993) in energy transport and the use of improved equation of
state at high densities. The important improvements in macrophysics and
numerics included the use of improved progenitor structure, the inclusion
of convection, the use of improved neutrino transport algorithm
(multi-group, flux limited, full transport, diffusion) and the inclusion
of general relativistic effects.  Other authors believe that the correct
explosion mechanism can only be demonstrated through multidimension (2-D
or a full 3-D) calculations. In fact, the recent VLB radio observations
and HST observations of SN1987A suggest that SN1987A and perhaps many
SNeII explode aspherically and perhaps with jetting of their debris. The
natal kicks to new born neutron stars may also be a result of aspherical
explosion.  Numerical calculations of such aspherical explosions require
multi-D codes. Although such multi-D codes have been developed and applied
to study core collapse SNeII (e.g., Herant et al. 1994; Burrows, Hayes and 
Fryxell 1995; Mezzacappa et al. 1996; Janka and M\"uller 1996), they still
do not include all the relevant physics: None is a full 3-D, none
incorporates general relativity, none has correctly treated all known
neutrino processes in the core, none adequately handles transport in
either the angular or radial direction. 

Neutrinos alone, in 1-D codes, do not seem to be able to revive the
stalled shock.  A variety of hydrodynamic instabilities have been invoked
by theorists over the years to help explode supernovae.  Neutrino
driven instabilities between the neutrinospheres and the stalled shock are
generic feature of core-collapse supernovae (Bethe 1990; Herant, Benz and
Colgate 1992; Herant et al. 1994; Burrows, Hayes, and Fryxell 1995;  Janka
and M\"uller 1996; Mezzacappa et al 1996).  Though it is generally
accepted that pre-explosion cores of massive stars are hydrodynamically
unstable, the role of convective motions in driving supernova explosions
is not yet clear. 

Core overturn driven by negative entropy and lepton gradients during the
deleptonization and cooling of the protoneutron star may boost the
driving neutrino luminosities (Burrows 1987;  Keil, Janka and M\"uller
1996). Only after a full neutrino transport will be incorporated in
multi-dimensional calculations it will become clear whether neutrinos can 
drive supernova explosions.

\noindent 
{\bf e. Is There a Neutron Star?} The 12 seconds neutrino burst from
SN1987A suggests the formation of a neutron star, at least transiently.
Besides the neutrino burst there is no other evidence that the SN1987A
remnant contains a central neutron star. Independent observations have
failed to confirm reported observation (Middleditch) of a 2.1 ms optical
pulsar. At present, the emission observed from SN1987A is completely
accounted for by radioactive energy sources (mainly $^{44}$Ti with half
life of 78 years) in the debris, so the energy input from a pulsar or any
other source must be small. To have escaped detection, the central compact
object must have a luminosity less than a few hundred times that of the
sun and far less than that of the 943 years old pulsar in the Crab Nebula. 
However, the average column density of the expanding shell (assuming
spherical symmetry) is $\approx 20M_\odot/4\pi R^2\approx 0.5~g~cm^{-2}$
for an average expansion velocity around $2500 km~s^{-1}$. The debris now
is quite cold throughout (a few hundred K only) and probably blocks the
light from the central source for decades, or longer if the debris are
clumped and the source happens to lie behind a cloud.  However, the
expanding shell is not opaque to energetic gamma rays. 

\noindent
{\bf f.g. A Central Black Hole ? Double Bang?} It was suggested that late 
time accretion
(Brown, Bruenn and Wheeler 1992) may have induced collapse of the nascent
neutron star into a black hole. Such a scenario may lead to two neutrino
bursts (``double bang'') well separated in time, and may explain the Mont
Blanc early signal (Aglietta et al 1987). But the Mont blanc early signal
implies unrealisticly large binding energy release in the first bang which
has not been detected by the Kamiokande, IMB and Baksan detectors. The
fingerprint of a central stellar black hole are difficult to detect. 
There is a chance to ``detect'' the central black hole only
if it is orbitted by a close companion which survived the explosion. 

\noindent
{\bf h. Natal Kick and Gravitational Radiation}
Pulsar locations (e.g., Taylor and Cordes 1993) and proper motion data
(e.g., Harrison, Lyne and Anderson 1993) imply that radio pulsars are a
high-speed population. Mean three-dimensional galactic speeds of
450$\pm$90 km s$^{-1}$ have been estimated (Lyne and Lorimer 1994), with
measured transverse speeds of individual pulsars reaching up to $\sim$1500
km s$^{-1}$.  Impulsive mass loss in a spherical supernova explosion that
occurs in a binary can impart to the nascent neutron star a substantial
kick (Gott, Gunn, and Ostriker 1970). However, theoretical studies of
binary evolution through the supernova phase have difficulty reproducing
the observed velocity distributions (Fryer, Burrows and Benz 1997). This
implies that neutron stars receive an extra kick at birth.  Anisotropic
neutrino radiation (Chugai 1984; Woosley 1987) have been invoked to
accelerate neutron stars. A 1\% dipole asymmetry in the neutrino radiation
of a neutron star's binding energy is sufficient to accelerate it to
$\sim$300 km s$^{-1}$. Jetting of the ejecta along the polar axis and
imbalance between the momenta of the two opposite jets can also be the
origin of the natal kick. Aspherical explosion, perhaps even jetting of
the debris are already evident in the high resolution VLB radio images
(Gaensler et al 1997) and in the recent HST images (Pun et al 1997). 
Merger scenarios and non axisymmetric collapse can lead to very
significant gravitational wave emission at typical frequencies of $\sim
c/2\pi R\sim $ a few kilo Hertz. However a reliable estimate of the
gravitational wave signal (wave form and light curve) probably will have
to wait until the explosion mechanism becomes clearer. Perhaps the
Caltech-MIT Laser Interferometric Gravitational Wave Observatory (LIGO)
will detect gravitational wave signals from SNeII, before reliable
theoretical estimates become possible ? 

\section{Limits On Particle Properties and Interactions} 
Astrophysics and cosmology provide test grounds for the standard model of
particle physics and extensions of the model over distances, time scales
and other conditions not accessible to laboratory experiments.  Limits on
neutrino properties from SNeII (lifetime, mixing, decay modes magnetic
moments) were derived long before SN1987A. SN1987A provided a new test
ground which attracted the attention of many more physicists. New limits
were derived not only for standard particles and minimal extensions of the
standard model, but for all kinds of hypothetical particles and
interactions.  Here, I will limit my summary to standard particles and
well motivated minimal extensions of the standard model of particle
physics. I will quote only limits which were derived from observations and
general considerations and either do not depend on , or are insensitive to
the detailed modeling of SNeII.  I will focus mainly on improvements since
1988 of the limits on neutrino properties which were included in Table 1
of my talk at La Thuile one year after SN 1987A (Dar 1988).  Table I, to
my judgement, summarizes the most important limits. 

\noindent 
{\bf Mass Limits On Standard Neutrinos.} The travel time of relativistic
neutrinos of mass $m_\nu$ and energy $E_\nu$ from a distance $D$ to Earth
is given approximately by
\begin{equation} 
             t=(D/c)[1+(1/2)(m_\nu c^2/E_\nu)^2].  
\end{equation} 
The observed energies and the dispersion in arrival times of the neutrinos
from SN1987A were used to estimate upper limits on the mass of the
$\nu_e$. Although the limits are model dependent, they are not very
sensitive to the models. A limit of about $m_{\nu_e}< 15~eV$ was obtained
both from simple models and from more ``sophisticated'' models (which are
not necessarily more reliable). The particle data group (Barnett et al.
1966) do not quote a laboratory limit since ``unexplained effects have
resulted in significantly negative $m_{\nu_e}^2$ in the new precise
tritium beta decay experiments''. The cosmological bound (e.g.,
Cowsik 1977 ), $\Sigma m_\nu<94\Omega h^2~eV$, yields 
$m_\nu < 15~eV$ for stable neutrinos, for the currently best measured
values of the cosmological parameters, $\Omega\leq 0.3$ and $h\approx 0.7$.
This limit on $m_{\nu_e}$ may be improved by one order
of magnitude by the more sensitive detectors like Superkamiokande and SNO
if they will detect a thermal neutrino burst from a more distant SNeII or
a neutronization burst from a galactic SNeII (e.g., Dar 1988). 

If neutrinos are Dirac particles with nonzero mass they can flip their
helicity in collisions with neutrons in the protoneutron star. Right
(left) handed neutrinos (antineutrinos) which have no standard electroweak
interactions escape immediately and cool the hot protoneutron star (PNS). 
Since the neutrino helicity flip cross section is proportional to
$m_\nu^2$ (standard electroweak $Z^0$ exchange yields
$\sigma_{flip}\approx G_F^2m_\nu^2/\pi$), the observed cooling rate of
SN1987A was used to obtain the limit $m_\nu<15~keV$ for Dirac neutrinos
(Raffelt and Seckel 1988; Griffols and Masso 1990; Dar 1990). This limit
is much weaker than the cosmological limit, but applies also to unstable
neutrinos with $\tau>R_{PNS}/\gamma c> 10^{-7}s~!$ It is much stronger
than the laboratory limits, $m_{\nu_\mu}<170~keV$ and $m_{\nu_\tau}<
24~MeV$. 

The Supernova mass limits on muon and tau neutrinos may be improved by
about two to three orders of magnitude by future measurements of the time
structure of the neutrino bursts from galactic SNeII with neutrino
telescopes like SNO, Superkamiokande and HELLAZ which are sensitive to all
neutrino flavors (e.g., Dar 1988). 

Finally, much stronger neutrino mass limits can be obtained from SNeII
neutrino bursts if neutrinos are mixed and oscillate. 

\noindent
{\bf Neutrino Oscillations.} The number of events which were detected by
Kamiokande and IMB, their angular distribution, and their energy
distribution and the maximal binding energy release in gravitational core
collapse suggest that they are mostly $\bar\nu_e p\rightarrow ne^+$
events. However, if the $\bar\nu_e$ were obtained by neutrino oscillation
from $\bar\nu_\mu$'s or $\bar\nu_\tau$'s into $\bar\nu_e$'s in vacuum with
a large mixing angle, their temperature should have been much higher,
$T\sim 7~MeV$. The observed temperature $T\leq 4~MeV$ of the Kamiokande
and IMB 19 events practically excludes the large angle vacuum oscillation
solution to the solar neutrino problem. 

\noindent
{\bf Neutrino Lifetime.}
If the neutrinos which were detected are those that were emitted
by SN1987A (no mixings, no ``conspiracy schemes''), 
then their mean life time must satisfy $\gamma\tau(\nu_e)
>D/c\approx 5\times 10^{12}s$, since they arrived with the expected
number (see, e.g., Bahcall, Dar and Piran 1987).

\noindent
{\bf Axion Mass.}
Various extensions of the standard model predict the existence
of light neutral pseudoscalars, like the axion proposed by Peccei and 
Quinn (PQ) in order to solve the problem of CP conservation in strong 
interactions. The original axion associated with the breaking of 
the PQ symmetry at the weak scale ($f_w$) is excluded 
experimentally, but not the invisible axion if the breaking scale 
is much larger ($f_a\gg f_w~~ m_a\sim f_a$). The axion lifetime is 
very long but a strong magnetic field can enhance its  
decay via $a\rightarrow \gamma_v\gamma$. Limits on the mass of the  
invisible axion were derived from laboratory experiments, astrophysics and 
cosmology. The absence of a $\gamma$ ray signal from SN1987A has
limited its mass to the narrow window $10^{-6} eV\leq m_a
\leq 10^{-3} eV$ (e.g., Raffelt 1990).   

{\bf What to Expect?} Detectors like Superkamiokande and SNO will allow
for the first time good energy, time and flavor spectroscopy of {\bf both}
the neutronization burst and the thermal burst from galactic SNeII. In
particular, the early phase of core collapse that precedes SNeII is better
understood than the explosion. Neutrinos from this phase are emitted
mainly due to $e^-$ captures on free protons and iron group nuclei. Up to
core densities of $\simeq 3 \times 10^{11}~g~cm^{-3}$ (neutrino trapping
density) these $\nu_e$ escape freely from the overlying stellar matter
without any interaction that changes their energy. The total number of
neutrinos emitted from a 1.4 $M_\odot$ stellar core as it evolves from a
initial density of $\sim 4 \times 10^9 g~cm^{-3}$ to a neutrino-trapping
density of $\simeq 3 \times 10^{11} g~cm^{-3}$ is $\simeq 10^{56}$.  The
duration of this $\nu_e$ burst is a few ms. The charge-current and
neutral-current reactions, $\nu_eD\rightarrow ppe^-$ and $\nu_xD\rightarrow
pn\nu_x$, respectively, on deuterium nuclei in SNO and the $\nu_x
e\rightarrow \nu_xe'$ scattering in the more massive Superkamiokande water
detector, can be used to detect the neutronization burst from galactic
SNeII, to identify its flavor content and measure the neutrino energies .
These may yield new important information on the physical and the nuclear
configuration of the collapsing stellar core and on neutrino properties,
in particular on neutrino masses, flavor mixing and matter oscillations
(Dar 1988). 

\section{More To Expect}
The story of SN1987A is not over yet. For astrophysicists
perhaps the most exciting future developments will be the collision of the
debris with the circumstellar gas and rings which will shed more light on
the nature of the explosion and on the history of the progenitor before
its supernova phase, the emergence of a neutron star and the birth of a
pulsar: 

\noindent 
{\bf Future Fireworks.} The blast wave from SN1987A will strike the
inner ring some six to ten years from now (Chevalier and Dwarkadas 1995;
Borkowski, Blondin and McCray 1997) and the ring is predicted to brighten
by a factor $\sim 10^3$ in all bands of the electromagnetic spectrum.
Shock acceleration will probably begin to produce relativistic particles
and $\gamma$ ray emission from the inner ring. 

\noindent
{\bf Shining the Past.} When the blast wave will continue to propogate
into the interstellar medium it will lit more rings and shells which may
have been ejected by the progenitor in its presupernova phase. 

\noindent
{\bf Neutron Star Emergence and Pulsar Birth.} The debris will
first become transparent to $\gamma$-rays and X-rays.  If the hot neutron
star is there, it will glow in thermal X-rays. If it has begun pulsed
emission over the whole electromagnetic spectrum and if we happen to lie
within the pulsar beaming cones, we will start to see pulsed emission of
radio waves, X-rays, and perhaps $\gamma$ rays. It will take much longer
(half a century or more) before the debris will become transparent to
optical photons. 

\section{Concluding Remarks}
Perhaps the most important consequences of SN1987A are the 
birth of extrasolar neutrino astronomy, the 
construction of galactic and extragalactic neutrino telescopes and the 
push to the construction of gravitational wave detectors. All these 
will help solve some of the most interesting puzzles in astronomy and 
test interactions and particle properties over physical domains
not accessible to laboratory experiments.  

\noindent
{\bf Acknowledgement}: The author would like to thank M. Greco and
G. Belletini for their generousity and for their friendship which 
has been extended to him over many years.  

\newpage

\pagestyle{empty}
\setlength{\evensidemargin}{-1cm}
\setlength{\oddsidemargin}{-1cm}
\textwidth = 16.5cm
\baselineskip 22pt plus 2pt
 
{\bf Table I:} Expected limits on neutrino properties 
from nearby SNeII compared with the 
corresponding limits from 
terrestrial experiments, from SN 1987A and from cosmology.

\begin{tabular}{|l|l|l|l|l|}\hline
{\bf Property} & {\bf Terrestrial Exp} & {\bf SN1987A} & {\bf Nearby SNeII} &
{\bf Cosmology} \\
               &                       &         &  \ \ ({\bf Expected}) &
\\ \hline
Masses &   &  & &     \\
$m_{\nu_e}$ & \ \ \ - & $<$15 eV & $<$1 eV & $<$15 eV \\
$m_{\nu_{\mu}}$ & 170 keV & $<$15~keV~(if~Dirac) & $<$100 eV & $<$15 eV \\
$m_{\nu_{\tau}}$ & 24 MeV & $<$15~keV (if~Dirac)& $<$100 eV & $<$15 eV \\ 
\hline Lifetime & (Atmospheric $\nu$'s) &  &  & \\
$\gamma \tau(\nu_e)$ & $>4 \times 10^{-2}$ s & $>5 \times 10^{12}$
s & $>10^{14}$ s & $>10^3$ s \\
$\gamma \tau(\nu_\mu)$ & $>4 \times 10^{-2}$ s & \ \ \ - &
$>10^{12}$ s & $>10^3$ s \\
$\gamma \tau(\nu_{\tau})$ & \ \ \ - & \ \ \ - & $>10^{12}$ s & $>10^3$ s
\\ \hline
Mixing  & Excluded Region & & & \\
$< \nu_e|\nu_x>$ & 
$\left\{ \matrix{\Delta m^2 > 0.1 eV^2 \cr
sin^2 2\theta>0.01} \right .$ 
& 
$ \matrix{ Large~Angle \cr
  Mixing~Excluded}$ 
&   
$ \matrix{ Small~Angles \cr
  Also~Excluded?}$ 
&    \\ \hline
Electric Charge & & & & \\
$q(\nu_e)$ & $<10^{-13}e$ & $<2 \times 10^{17}e$ & $<1 \times 10^{-18}e$ & \\
$q(\nu_\mu)$ & $<10^{-6}e$ & \ \ \ - & $<2 \times 10^{-17}e$ & \\
$q(\nu_\tau)$ & $<10^{-2}e$ & \ \ \ - & $<2 \times 10^{-17}e$ & \\ \hline
Magnetic Moment & & & & \\
$\mu (\nu_e)$ & $1.8 \times 10^{-10}\mu_B$ & $<10^{-12}\mu_B$ &
$<10^{-14}\mu_B$ & \\
$\mu(\nu_\mu)$ & $7.4 \times 10^{-10}\mu_B$ & $<10^{-12}\mu_B$ &
$<10^{-14}\mu_B$ & \\
$\mu(\nu_\tau)$ & $5.4 \times 10^{-7}\mu_B$ & $<10^{-14}\mu_B$ &
$<10^{-14}\mu_B$ & \\ \hline
Radiative Decay & & & & \\
$B^{-1}_\gamma \tau_\nu/m_\nu$ &
$>20$ s/eV & $>2 \times 10^{16}$ s/eV & $>10^{17}$ s/eV &
  \\
  &     &        $(m_\nu > 20$ eV) & $(m_\nu > 20$ eV) & \\ \hline
$\nu$ Flavors & \ \ \ 3 & $\leq 5$ & \ \ \ 3 & \ \ 3 \\ \hline
\end{tabular}
\bigskip

\noindent
Fig.1: An image of the triple ring structure around the remnant of SN1987A
taken in early 1997 by the Wide Field and Planetary Camera 2 of the Hubble 
Space Telescope.

\end{document}